\newif\ifieee  
\ieeetrue
\newif\ifieeefinal
\ieeefinaltrue
\newif\ifedit
\editfalse
\newif\ifwarn   
\warntrue
\newif\ifwarnshort  
\warnshortfalse
\newif\iftodo   
\todotrue
\newif\iftodoshort  
\todoshortfalse
\newif\ifmore   
\moretrue
\newif\ifmoreshort  
\moreshorttrue
\ifieee
  \ifieeefinal
    \documentclass[a4paper,twoside,twocolumn]{IEEEtran}
  \else
    \documentclass[a4paper,onecolumn,draftcls]{IEEEtran}
  \fi 
\else
  \documentclass[11pt,fleqn]{article}
\fi 
\ifieee
\else
\fi
\ifedit
  \usepackage[german,english]{babel}   
\fi
\usepackage{amsmath}
\usepackage{amssymb}
\ifieee
  \usepackage{theorem}
\else
  \usepackage{amsthm}
\fi
\usepackage{epsfig}
\usepackage{xspace}
\usepackage{algorithm}
\usepackage{algorithmic}
\usepackage{nicefrac}
\ifedit
  \usepackage{mycolor}
  \usepackage{manfnt}  
  \usepackage{marvosym}  
  \DeclareMathSymbol{\myRightarrow}{3}{symbols}{41} 
\fi
\newcommand{\emptypage}%
{
  \newpage
  \vspace*{10cm}
  \pagebreak
}
\ifieee

\else

\fi
\newcommand{\more}[2]     
{\ifmore
  \ifmoreshort
    \reversemarginpar\marginpar{{\sf\footnotesize More?}\,
      \begin{minipage}[c]{0.05\linewidth}\tiny #1 \end{minipage}}
  \else
    \par
    \reversemarginpar\marginpar{{\sf\footnotesize More?}\,
      \begin{minipage}[c]{0.05\linewidth}\tiny #1 \end{minipage}}
    \fbox{\begin{minipage}[t]{0.95\linewidth}\selectlanguage{german}
       #2
    \end{minipage}
    }
    \par
  \fi 
\fi}
\ifieee
  \theoremstyle{plain}
  \theoremheaderfont{\bf}
  {\theorembodyfont{\it}
  \newtheorem{theorem}{Theorem}[section] 
  \newtheorem{proposition}[theorem]{Proposition}
  \newtheorem{lemma}[theorem]{Lemma}
  \newtheorem{corollary}[theorem]{Corollary}
  \newtheorem{conclusion}[theorem]{Conclusion}
  }
  {\theorembodyfont{\rm}
  \newtheorem{definition}[theorem]{Definition}
  }
  {\theorembodyfont{\rm}
  \newtheorem{remark}[theorem]{Remark}
  \newtheorem{note}[theorem]{Note}
  \newtheorem{cit}[theorem]{Citation}
  }
\else
  \theoremstyle{plain}
    
  \newtheorem*{theorem*}{Theorem}
  
  \newtheorem*{proposition*}{Proposition}
  
  \newtheorem*{lemma*}{Lemma}
  
  \newtheorem*{corollary*}{Corollary}
  
  \newtheorem*{conclusion*}{Conclusion}
  \theoremstyle{definition}
  
  \newtheorem*{definition*}{Definition}
  \theoremstyle{remark}
  
  \newtheorem*{remark*}{Remark}
  
  \newtheorem*{note*}{Note}
  
  \newtheorem*{cit*}{Citation}

\fi 
\ifieee\else
  
\fi
\newcommand{\mathcomma}{\:,\:\:}

\newcommand{\mapto}[2]{#1\longmapsto #2}
\usepackage{dsfont}
\newcommand{\CC}{\ensuremath{\mathds{C}}}   
\newcommand{\RR}{\ensuremath{\mathds{R}}}

\newcommand{\mdef}{:=}

\newcommand{\oneover}[1]{\frac{1}{#1}}

\newcommand{\comp}{\raisebox{0.4ex}{$\scriptscriptstyle\circ$}}

\newcommand{\fnorm}[1]{\lVert#1\rVert_{\scriptscriptstyle\text{F}}}
\newcommand{\bigfnorm}[1]{\left\lVert#1\right\rVert_{\scriptscriptstyle\text{F}}}

\newcommand{\Unity}{\ensuremath{\mathbf{1}}\xspace}
\newcommand{\Zero}{\ensuremath{\mathbf{0}}\xspace}

\newcommand{\textmatrix}[1]{
   \left(\begin{smallmatrix}#1\end{smallmatrix}\right)}

\newcommand{\equivalent}{\ensuremath{\quad\Longleftrightarrow\quad}}

\newcommand{\tends}{\ensuremath{\longrightarrow}}

\DeclareMathOperator{\diag}{diag}

\DeclareMathOperator{\dimr}{\dim_{\RR}}

\renewcommand{\mod}{\ensuremath{\,\text{mod}\,}}
\newcommand{\fft}{\ensuremath{\text{\sc fft}}\xspace}
\newcommand{\ifft}{\ensuremath{\text{\sc ifft}}\xspace}
\newcommand{\K}{\ensuremath{K}\xspace}
\renewcommand{\L}{\ensuremath{L}\xspace}
\newcommand{\T}{\ensuremath{T}\xspace}
\newcommand{\nt}{\ensuremath{n_{t}}\xspace}
\newcommand{\nr}{\ensuremath{n_{r}}\xspace}
\newcommand{\Vc}[1]{\ensuremath{V^{\CC}_{#1}}}  
\newcommand{\kbein}{\ensuremath{\textmatrix{\Unity\\\Zero}}}

\newcommand{\daggered}[1]{\ensuremath{#1^{\dagger}#1}}
\newcommand{\ext}[1]{\ensuremath{{#1}_{_{\text{mp}}}}}
\newcommand{\sym}{\mathrm{sym}}
\newcommand{\Phiext}{\ensuremath{\ext{\Phi}}}
\newcommand{\Psiext}{\ensuremath{\ext{\Psi}}}
\newcommand{\deltakc}{\ensuremath{\Delta}}
\newcommand{\deltakcext}{\ensuremath{\ext{\Delta}}}
\newcommand{\varrhokcF}{\ensuremath{\varrho_{_F}}}
\newcommand{\dkc}{\ensuremath{\mathrm{d}}\xspace}
\newcommand{\dkcF}{\ensuremath{\mathrm{d}_{_F}}\xspace}
\newcommand{\dkcT}{\ensuremath{\mathrm{d}_{_T}}\xspace}

\newcommand{\pkcF}{\ensuremath{\mathrm{p}_{_F}}\xspace}
\newcommand{\pkcT}{\ensuremath{\mathrm{p}_{_T}}\xspace}

\newcommand{\chkcF}{\ensuremath{\mathrm{ch}_{_F}}}

\newcommand{\divkcF}{\ensuremath{\mathcal{D}iv}_{_F}}

\title{Space Frequency Codes from Spherical Codes}
\author{
  Oliver Henkel\\
  Fraunhofer German-Sino Lab for Mobile Communications - MCI\\
  Einsteinufer 37, 10587 Berlin, Germany\\
  Email: {\tt henkel@hhi.fraunhofer.de}
     }
\date{}
\begin{document}
\maketitle
\begin{abstract}
A new design method for high rate, fully diverse ('spherical') space
frequency codes  
for MIMO-OFDM systems is proposed, which works for arbitrary numbers of
antennas and subcarriers. The construction exploits a differential
geometric connection between spherical codes and space time codes.
The former are well studied e.g. in the context of optimal sequence design
in CDMA systems, while the latter serve as basic building blocks for space
frequency codes.
In addition a decoding algorithm with moderate complexity is
presented. This is achieved by a lattice based construction of spherical
codes, which permits lattice decoding algorithms and thus 
offers a substantial reduction of complexity.
\end{abstract}
\section{Introduction}
In MIMO (Multiple Input Multiple Output) systems space time coding schemes
have been proven to be an appropriate tool to combat fading
holes of Rayleigh flat fading channels and exploit the spatial diversity 
gains. For frequency selective SISO (Single Input Single Output) channels
(due to multipath propagation)
OFDM (Orthogonal Frequency Division Multiplexing) provides an efficient
strategy to obtain a collection of mutually orthogonal flat fading channels
on each subcarrier. The number $\K$ of subcarriers in such a system is not a
free design parameter but 
predetermined by the bandwidth (which determines the sampling rate) and the
coherence length of the channel (which determines the maximal OFDM symbol
length, which is the maximum number of subcarriers). 
MIMO-OFDM is the combination of MIMO techniques to frequency selective
Rayleigh (multipath) channels. Space frequency coding aims to combine the
advantages of space time coding with the multipath diversity from each
channel tap.  

A corresponding code design 
for fully diverse space frequency codes $\mathcal{C}$ has been
proposed in \cite{boe.bor.pau}. In their work they separated the code
construction into 
the design of an appropriate transition matrix $A$ incorporating the multipath
diversity and the construction of a corresponding space time code
$\tilde{\mathcal{C}}$. This method has the advantage of splitting up
the design problem into two separated problems by keeping the multipath
echoes of the transmit signal mutually orthogonal. For these reasons this
work focuses on the design method \cite{boe.bor.pau}, although other approaches
have been investigated (e.g. \cite{liu.xin.gia}).

To obtain maximum rates by retaining full (spatial and
frequency) diversity order the 
block length $T$ of the 'inner' space time code has to be as large as possible
(i.e. $T=\K/\L$, whereas $\L$ denotes the number of taps). This requirement
rules out most of the space time code designs considered 
in the literature so far. Common research has been focused mainly on small
dimensional coding spaces (i.e. $\T=\nt$) whose elements consist of square
$\nt\times\nt$ matrices, whereas $\nt$ denotes the number of transmit
antennas. These designs correspond to the case $\K/\L=\nt$ only. To
attack the more general (space time) code design problem an inspection of
the maximum likelihood decision criterion reveals, that it is a good first
approach to consider code design as a packing problem. The corresponding
coding space turns out to be the complex Stiefel manifold of complex
$T\times\nt$ matrices with unit norm mutually complex
orthogonal column vectors. 

On the other hand the design of optimal (Welch bound equality WBE) signature
sequences in CDMA (Code Division Multiple Access) leads naturally to a
packing problem on spheres \cite{mas.mit,hea.str.pau}. Explicit
constructions of spherical packings can be found on N.J.A. Sloane's web
page\footnote{
{\tt www.research.att.com/\raisebox{0.25ex}{$\scriptstyle\sim$}njas/packings/}}.

The first achievement of this work is to employ differential geometric
methods to establish a connection between spherical packings/codes and
(general) space time packings/codes, such that the rich variety of already
known spherical 
codes can be transformed to space time codes in various dimensions. This
mapping preserves 
(geodesic) distances, and therefore the performance at the receiver can be
controlled (up to a certain degree). The so constructed codes are called
\emph{spherical space time codes} and the resulting space frequency codes
(obtained from the procedure described above) are named \emph{spherical
  space frequency codes} from now on.

The second part of this work is concerned with decoding strategies. The
need for high rate code designs demands advanced decoding algorithms. Even
moderate numbers $\K$ of subcarriers need huge codebooks of size $2^{\K R}$
in order to retain a specified rate $R$, making brute force maximum
likelihood decoding 
impossible for practical applications. Based on wrapped spherical codes
\cite{ham.zeg} which are decodable in terms of lattice algorithms, the
connection to space time codes will be exploited to do the same for space
time and space frequency codes. This preferred decoding strategy prohibits
high rate spherical code classes obtained elsewhere, such as Sloane's
dense packings or packings obtained from frame theory
(e.g. \cite{str.hea}). 

On the other hand wrapped spherical
coding offers a wide range of lattice based constructions which can be
adapted to arbitrary numbers of antennas and subcarriers. Moreover for
each lattice 
of the appropriate dimension a target minimal distance can be prescribed,
leading to a corresponding wrapped spherical code which can be processed
further to obtain space time and space frequency codes. In this way whole
(fully diverse) families of codes can be constructed for each prescribed
number of antennas and subcarriers, whereas the rate can be adapted by
choosing appropriate minimal distances. 

The text is structured as follows. In the next two sections the channel
model and the basic code design criteria will be derived. Section 
\ref{s.stc2sfc} is devoted to the design of space frequency codes, which has
been made popular in \cite{boe.bor.pau}. Section \ref{s.s2stc} sketches
the differential geometric connection between spherical codes and space
time codes and describes the corresponding mapping.
In section \ref{s.decoding} the second part of this work starts. It
contains the encoding and decoding procedure (based on
\cite{ham.zeg} for the spherical part) for the proposed spherical space
frequency codes, 
which are simulated in section \ref{s.simulations}. Finally in the last
section the results are summarized for a quick reference.
\section{Channel model}\label{s.channel}
Let us adopt the following notations:
$\hat{x}=\fft(x)$, $x'=\ifft(x)=\fft^{\dagger}(x)$, 
$\bar{\cdot}$ denotes complex conjugation, $\cdot^t$ the transpose, 
$\cdot^{\dagger}$ the hermitian transpose, $\Unity$ denotes the identity
matrix and the all-zero matrix is denoted as $\Zero$, whereas their
dimensions will be clear from the context, if not indicated by some subscripts.
Using the common (complex conjugated) {\K}-th root of unity 
$\omega=e^{-\frac{2\pi\imath}{\K}}$
the (\K-point) (I)FFT matrix reads 
$(\text{\sc i})\fft=[(\text{\sc i})\fft_{kk'}]_{k,k'=0}^{\K-1}$ with components
$\fft_{kk'}=\oneover{\sqrt{\K}}\omega^{kk'}$, resp.
$\ifft_{kk'}=\oneover{\sqrt{\K}}\bar{\omega}^{k'k}$.
The frequently used operations of (frequency) modulation 
$\mu=\diag(\omega^k)_{k=0}^{\K-1}$ and (time) translation 
$\tau=[e_{(k+1)\mod \K}]_{k=0}^{\K-1}$ 
on $\CC^{\K}$ ($e_k$ denoting the $k$-th standard basis vector) transform
into each other by  
$\mu=\fft\comp\tau\comp\ifft$, resp.
$\tau=\ifft\comp\mu\comp\fft$.

The channel model assumed here is the frequency selective Rayleigh fading
MIMO-OFDM channel with \K subcarriers, \L taps
$H_l\in\CC^{\nt\times\nr}\sim\text{i.i.d. } {\cal CN}(0,\oneover{\L})$,
and \nt, (resp. \nr) transmit (resp. receive) antennas. 
Before transmission a \K point IFFT is applied to each $\K\times\nt$
MIMO-OFDM code symbol 
$C=(c_0^t,\dots,c_{\K-1}^t)^t$, 
which is then prepended by the cyclic prefix of
length \L. The
receiver applies a \K point FFT and removes the cyclic prefix
before (maximum likelihood) decoding. 
Then, the transmission equation reads in the frequency domain
\begin{equation}\label{e.f-channel}
   \tilde{c}_k 
     = \sqrt{\rho}\, c_k \hat{H}_k + n_k
   \mathcomma
   k=0,\dots,\K-1
\end{equation}
whereas 
\begin{equation}
\hat{H}_k = \sum_{l=0}^{\L-1} H_l e^{-\frac{2\pi\imath}{\K}kl}
\end{equation}
and 
$n_k\in\CC^{1\times\nr}$ (noise), with both $\hat{H}_k$ and $n_k$ having
i.i.d. $\mathcal{CN}(0,1)$ distributed components for all $k=0,\dots,\K-1$.
It is assumed that the transmitter knows only the channel statistic while the
receiver perfectly knows the channel coefficients. 
The rows $c_k$ of $C$ are assumed to have unit norm, such that 
$\rho$ is the SNR (signal to noise ratio) per subcarrier at each receive
antenna.   
\section{Space frequency code design criteria}\label{s.design-criteria}
This section describes the code design criteria developed in
\cite{boe.bor.pau} with emphasis on the underlying geometrical
framework. Let us express the transmitted symbols $C$ as 
$C \mdef \sqrt{\rho\frac{\K}{\nt}}\, \Phi$,
assuming $\Phi$ to be an element of the complex Stiefel manifold 
\begin{equation}
   \Vc{\nt,\K}=\{\Phi\in\CC^{\K\times\nt}\,|\,\Phi^{\dagger}\Phi=\Unity\}
\end{equation}
which is the basic coding space from now on, compare
\cite{hoc.mar.1},\cite{boe.bor.pau}. 
A space frequency code is given by a discrete set 
$\mathcal{C} \subset \Vc{\nt,\K}$ and provided the received signal
$\tilde{C}=\sqrt{\rho\frac{\K}{\nt}}\,\Psi H+N$
the maximum likelihood decision reads (see \cite{hoc.mar.2})
\begin{equation}\label{e.ml-detection}
   \Phi_{\text{ML}} = \arg\min_{\forall_{\Phi\in{\cal C}}}
      \bigfnorm{\tilde{C}-\sqrt{\rho\frac{\K}{\nt}}\,\Phi H}
\end{equation}

In the sequel let us adopt the notation that in order to emphasize the 
space {\bfseries F}requency (code) or space {\bfseries T}ime (code)
interpretation of a given quantity 
subscripts $\cdot_{_F}$, respectively $\cdot_{_T}$ will be added.
Further, for the difference symbol $\deltakc=\Phi-\Psi$ of two codesymbols 
we define the '{\bfseries m}ulti{\bfseries p}ath extension'
$\deltakcext = (\deltakc,\mu\deltakc,\dots,\mu^{\L-1}\deltakc)$. Then the
Chernov bound for the probability of mistaking $\Phi$ for $\Psi$ is given
by (compare \cite{boe.bor.pau})
 \begin{equation}
    \chkcF = \left(\prod_{i=1}^{\nt\L} 
                 \left[
                   1 + \varrhokcF \sigma_i^2(\deltakcext)
                 \right]\right)^{-\nr}
 \end{equation}
whereas $\varrhokcF\mdef \oneover{4}\rho\frac{\K}{\nt\L}$ and 
$\sigma(M)=(\sigma_i(M))$ generically denotes the vector of singular values of the
matrix $M$ in decreasing order. 
Expanding the product leads to the diversity, resp. diversity sum \dkcF and
product \pkcF
\begin{equation}\label{e.divkcF}\begin{split}
   \divkcF \mdef \prod_{i=1}^{\nt\L}
     \left[
       1 + \varrhokcF\sigma_i^2(\deltakcext)
     \right]
     = \sum_{i=0}^{\nt\L}s_i\varrhokcF^i 
\end{split}\end{equation}
whereas ($\sym_i$ denotes the $i$-th elementary symmetric polynomial)
$s_i\mdef
 \sym_i(\sigma_1^2(\deltakcext),\dots,\sigma_{\nt\L}^2(\deltakcext))$ 
and
\begin{gather}
   \label{e.dkcF}
   \dkcF^2 \mdef s_1
           = \fnorm{\deltakcext}^2 
           = \L\,\fnorm{\deltakc}^2
           = \L\,\dkcT^2\\   
   \begin{split}
   \label{e.pkcF}
   \pkcF^2 \mdef s_{\nt\L} =\det(\daggered{\deltakcext})
           \le (\pkcT^2)^{\L}
   \end{split}
\end{gather}
with $\dkcT\mdef\fnorm{\deltakc}$, $\pkcT\mdef\sqrt{\det{\daggered{\deltakc}}}$
denoting the corresponding diversity sum and diversity product
for space-time codes. Note, that due to $\dkcF^2=\L\dkcT^2$ the maximum 
likelihood detection for space frequency codes is
essentially the same than for space time codes. Further, equality in
(Fischer's inequality)
\eqref{e.pkcF} holds, if 
$\deltakc^{\dagger}\mu^l\deltakc=0\,\forall_{l=1,\dots,\L-1}$, provided 
$\pkcT\neq0$. 
Thus a sufficient condition for the maximization of diversity
(sum/product) is
\begin{subequations}\label{e.max-div-kc}
\begin{gather}
   \label{e.dkcT}
   \max:\: \dkcT^2 = \fnorm{\Phi-\Psi}^2 \overset{\text{!}}{>}0 \\
   \label{e.pkcT}
   \max:\: \pkcT^2 = \det{\daggered{\deltakc}} \overset{\text{!}}{>}0 \\
   \label{e.max-div-order-kc}
   \forall_{l=1,\dots,\L-1}
      \big(
        \deltakc^{\dagger}\mu^l\deltakc = 0 
        \equivalent
        \deltakc'^{\dagger}\tau^l\deltakc' = 0         
      \big)
\end{gather}
\end{subequations}
performed over all pairs $\Phi,\Psi$
of code symbols satisfying $\Phiext,\Psiext\in\Vc{\nt\L,\K}$.
So, while the first two conditions are borrowed from the theory of space
time codes, the third set of conditions is a serious constraint: The
difference vector $\deltakc$ of two code symbols $\Phi,\Psi$ has to be
perpendicular to its row-wise $l$-th order modulated (resp. its
$l$-th order down-shifted) version for $l=1,\dots,\L-1$, reflecting the
strong correlation between each $l$-th channel realization in the frequency
domain. The demanded orthogonality forces the best possible (frequency)
diversity induced by the multipath delay spread. 

The homogeneity of the diversity \eqref{e.divkcF} permits the
interpretation of \eqref{e.max-div-kc} as a constraint packing
problem on the Stiefel manifold $\Vc{\nt\L,\K}$: Fully diverse space
frequency coding becomes a packing problem in $\Vc{\nt,\K}$ (due to
\eqref{e.dkcT}) subject to the 'space time' constraint \eqref{e.pkcT} and
the 'space frequency' constraint \eqref{e.max-div-order-kc}.
For space time codes the corresponding interpretation has been pointed out
in \cite{hen-2}.  
\section{Space frequency codes from space time codes}\label{s.stc2sfc}
As has been pointed out in \cite{boe.bor.pau}, the design of fully
diverse space frequency codes splits into two independent parts, reducing
the overall complexity of the problem considerably. First construct a 
space time code $\tilde{\mathcal{C}}'\subset\Vc{\nt,T}$ (subject to
the first two conditions in \eqref{e.max-div-kc}) for some
$T$. Then find some matrix $A'$ satisfying 
\begin{equation}
   \ext{A}'=(A',\tau A',\dots,\tau^{\L-1} A')\in\Vc{\T\L,\K}
\end{equation}
The final space frequency code fulfilling
\eqref{e.max-div-order-kc} is then given as \cite{boe.bor.pau}
\begin{equation}\label{e.factorisation}
   \mathcal{C}=\fft (A' \tilde{\mathcal{C}}')
\end{equation}
From this construction the rate $R$ of the space frequency code (neglecting
the cyclic prefix) and the rate $\tilde{R}$ of the internal space time
code are related by
\begin{equation}
   R=\frac{\T}{\K}\tilde{R}
\end{equation}
from which it is obvious that high rate codes require large block lengths
\T. On the other hand to retain the maximal diversity order $\L\nt\nr$ of
the space frequency code $\T\le\K/\L$ is demanded. Therefore the code
design aims at code constructions satisfying
\begin{equation}\label{e.tkl}
 T=\frac{\K}{\L}
\end{equation} 
and we concentrate on this case in the sequel.

As a further notational convention we introduce the tilde $\tilde{\cdot}$
for quantities corresponding to the 'inner' space time code 
$\tilde{\mathcal{C}}'$. Recall that $\mathcal{C}$ in
\eqref{e.factorisation} has full (frequency and spatial) 
diversity, if $\tilde{\mathcal{C}}'$ is a fully diverse space time code
and for large block lengths this is a complex code design problem.
Therefore the strategy in this work is to concentrate on the simpler first order
design criteria \eqref{e.dkcT} in the next section and develop some
adjustments for full diversity later on (section \ref{s.decoding}), compare
the remark in the last 
paragraph in section \ref{s.design-criteria}.   
Due to the factorization \eqref{e.factorisation} the goal investigated in the
next section is to find dense packings in the metric space
$(\Vc{\nt,\T},\tilde{\dkc}_T)$.

Let us close this section by some analysis of the frequency diversity
generating matrix 
$A'$. Random constructions have been proposed in \cite{boe.bor.pau}, but
here we present a deterministic construction which in addition is especially
designed to benefit from a low crest-factor: Set 
\begin{equation}\label{e.lambda-biunimodular}
\lambda=\left(\exp\left(\frac{2\pi\imath k^2}{\K}\right)_{k=0}^{\K-1}\right)
\end{equation}
(it is well-known that this sequence belongs to the set of so-called
bi-unimodolar sequences, i.e their FFT is as well unimodular). Then the matrix 
$U(c)=(u_0,\dots,u_{\K-1})\mdef(c,\tau c,\dots,\tau^{\K-1} c)$
with $c=\fft^{\dagger}\big(\oneover{\sqrt{\K}}\lambda\big)$ is a circulant
matrix which is also unitary. Let us define
\begin{equation}\label{e.standardA}
   A' \mdef (u_0,u_{\L},\dots,u_{(\T-1)\L})
\end{equation}
then  
$\tau^{l}A'=(u_l,u_{\L+l},\dots,u_{(\T-1)\L+l})$ and
$\ext{A}'^{\dagger}\ext{A}'=\Unity$ as desired.
This choice of $\lambda$
distributes the transmit energy equally in the time domain (compare
\cite{stanczak_05_sig} and references therein for an account to the
construction of sequences with zero autocorrelation sidelobes and zero
crosscorrelation). In general this construction scheme provides a real $\K$
dimensional 
parameter space $\lambda_k=e^{\imath\phi_k}$, $0\le\phi_k\le 2\pi$,
$k=0,\dots,\K-1$ for $A'$. 
\section{Space time packings from spherical codes}\label{s.s2stc}
The Stiefel manifold is canonically a homogeneous space 
$\big(U(\K) \overset{\pi}{\tends} \Vc{\nt,\K};\, U(\K-\nt)\big)$:
The canonical left multiplication  
of $\nt$-frames in $\CC^{\K}$ by unitary $\K \times \K$ matrices 
transforms each pair of $\nt$-frames isometrically (with respect to the
metric induced from the unitary group) into each other. Thus the action of the
unitary group $U(\K)$ on $\Vc{\nt,\K}$ is transitive and establishes the 
canonical diffeomorphism
\begin{equation}
   \Vc{\nt,\K} \cong U(\K) \left/ \textmatrix{\Unity & \Zero\\ \Zero & U(\K-\nt)}
                \right.
\end{equation}
Homogeneity is
the natural generalization of spherical symmetry: At any point the manifold
'looks the same' i.e. for any pair $(\Phi,\Psi)$ of points in $\Vc{\nt,T}$
there exists an isometry (with respect to the Riemannian metric structure)
which takes $\Phi$ into $\Psi$. In contrast to spheres the sectional
curvature is no longer constant, but remains non-negative and bounded from
above. Since positive curvature contracts distances, the homogeneity
$\Vc{\nt,T}$ suggests certain (with sectional curvature equal to the upper
bound just mentioned) $D$-dimensional spheres as
simple comparison spaces, 
\begin{equation}\label{e.D}
   D\mdef\dimr\Vc{\nt,T}=\nt(2T-\nt)
\end{equation}
Any code 
on such a sphere can be transformed into a code on the Stiefel manifold 
without shortening its geodesic distances or volumes (compare \cite{hen-1}
for details). 

In practice the following has been done. By homogeneity in both spaces one
'base' point has been fixed, e.g. the north pole $N$ of the sphere and the
point 
$\kbein$ in the Stiefel manifold. All points on the respective manifold can
be equivalently addressed in the corresponding tangent space (with respect
to the fixed base point) by utilizing the so-called exponential map, which
turns radial tangent lines isometrically into geodesics emanating from the
base point. Representing the Stiefel manifold as a $U(T)$-homogeneous
space and similarly the sphere as 
\begin{equation}
   S^D\cong O(D+1)/O(D)
\end{equation} 
($O(D)$ denoting the group of $D \times D$ orthogonal matrices), the
exponential map $\exp$ 
coincides with the matrix exponential 
$\exp:\mapto{A}{\sum_k A^k/k!}$. Restricting $\exp$ to the ball with radius
$\pi$, $\exp$ becomes a diffeomorphism onto the corresponding manifolds up
to a set of measure zero. Neglecting this sparse set, we have now a
representation of the sphere and the Stiefel manifold in terms of their
tangent space at the chosen base point.
Choosing some orthonormal base $\{Y_i\}$, resp. $\{X_i\}$ in each of
tangent spaces induces a 
canonical mapping from (the tangent space of) the sphere to the (tangent
space of the) Stiefel manifold by 
\begin{equation}\label{e.tangisom}
   \mapto{\sum_k v^k Y_k}{\sum_k v^k X_k}
\end{equation}
This mapping is an isometry of tangent spaces and due to curvature
non-radial distances will be contracted to an extend which is determined by
the sectional curvature upper bound in the Stiefel manifold. Moreover the
geodesic distance is only an approximation to the 'chordal' distance
$\tilde{\dkc}_T$ of the receiver. But it can be shown, that both
distance measures are (locally) equivalent \cite{hen-1}. In summary,
spherical codes (for example from MWBE sequences in CDMA systems
\cite{hea.str.pau}) can be transformed into space time codes, whereas
spherical 
minimal distances are possibly contracted, but only to an extend, which is
determined by the (bounded) curvature of the Stiefel manifold and the local
equivalence of the distance measures.  
\section{Lattice spherical space frequency codes and diversity rotation}
\label{s.decoding}
The construction of high rate codes in systems with large block lengths or
large number of subcarriers suffers from the increasing (brute force)
decoding complexity, which becomes intractable soon. To fight this,
additional structures in the code design need to be exploited. One way of
embedding (algebraic) structure is provided by the design algorithm for
wrapped spherical codes \cite{ham.zeg}: This algorithm uses a clever and
computationally simple one-to-one, distance expanding mapping from
$\RR^{D}$ to the sphere $S^{D}\subset\RR^{D+1}$ 
to map lattices on (each hemi-)sphere. Having this map and its inverse
at hand we can describe the overall encoding and decoding procedure by 
Algorithm \ref{alg.endecoding}.
\begin{algorithm}
\caption{Encoding ($1\rightarrow8$) / Decoding ($8\rightarrow4$)}
\label{alg.endecoding}
\begin{algorithmic}[1]
\REQUIRE \begin{minipage}[t]{0.9\linewidth}
$\K$ no. subcarriers\\
$\nt$ no. transmit antennas\\
$\L$ no. channel taps\\
$\dkc_S$ target (spherical) minimal distance
\end{minipage}
\ENSURE $\mathcal{C}$ space frequency code
\STATE $D \leftarrow \nt(2\T-\nt)$, $T \leftarrow \K/\L$
\STATE Choose some lattice $\Lambda\subset\RR^{D}$
\STATE Scale lattice generator to meet design distance $\dkc_S$
\STATE Apply wrapping algorithm in \cite{ham.zeg} to obtain a spherical code
   $(\mathcal{C}_S,\dkc_S)$
\STATE Project $\mathcal{C}_S$ to the tangent space $T_NS^{D}$ at the north
   pole (orthogonal projection 'upwards', then rescaling to length equal to
   the arclength between the considered point of $\mathcal{C}_S$ and the
   north pole $N$)
\STATE Map tangential code points to $T_{\kbein}\Vc{\nt,T}\cong T_NS^{D}$ via
   \eqref{e.tangisom}; this yields a tangential space time code
   $\tilde{\mathcal{C}}'_{\text{tang}}$
\STATE Apply the matrix $\log$ function to get 
   $\tilde{\mathcal{C}}'=\log\tilde{\mathcal{C}}'_{\text{tang}}
    \subset\Vc{\nt,T}$
 \STATE Apply \eqref{e.factorisation} with \eqref{e.standardA},
 \eqref{e.lambda-biunimodular} on $\tilde{\mathcal{C}}'$, which defines
 $\mathcal{C}$ 
\end{algorithmic}
\end{algorithm}
Note that the lattice encoding scheme permit large families of
code constructions based on (apart from the standard cubic lattices $Z_n$)
the root lattices (and their duals) $A_n$ ($A_n^*$), $D_n$ ($D_n^*$)
in various dimensions $n$, the exceptional lattices $E_{6,7,8}$, the  
Coxeter-Todd lattice $K_{12}$, the Barnes-Wall lattice $\Lambda_{16}$, and
the Leech lattice $\Lambda_{24}$, compare \cite{con.slo} for details. The
choice of the design distance $\dkc_S$ trades minimal distance for
rate.
For decoding just reverse the encoding algorithm until step 4 and apply some
lattice decoding afterwards. Note that this procedure is suboptimal
in general due to some points possibly getting lost in step 4 of the
encoding algorithm (compare \cite{ham}). But in the high rate regime it is
expected, that this possible performance loss becomes
negligible. Furthermore some efforts have been made to compensate the
'buffer regions' (\cite{ham}) in part: These regions originally placed as
separating annuli on the sphere to guarantee the designed minimal distance
$\dkc_S$ have been reinspected whether formerly rejected 'buffered' code
points lie at a distance at least $\dkc_S$ from all other code points on
the sphere. 

A common problem of the proposed lattice based construction is the inherent
rigidity of lattices. Although there is some control over the minimal distances,
there is no direct control on the diversity order, i.e. the diversity
product $\pkcF$. To overcome the possibility of diversity loss several
rotated versions of the lattices have been produced as a preprocessing step
of the encoding algorithm. Of course this procedure does not affect the
decoding performance. As the
axis of rotation the 'diagonal' $e$ defined as the sum of all basis vectors
of the respective lattice has been taken. Denote by $W_e$ the
orthogonal matrix with $e_1=W_e e$ whereas
$e_1=(1,0,\dots,0)^t$. 
Set $\Zero=(0,\dots,0)^t\in\RR^{D-1}$,
$R_1=\textmatrix{
              1     & \Zero^t\\
              \Zero & R^{\perp}
     }
$
and define $R^{\perp}=\exp(\alpha X)$,
$\alpha\in\RR$, with
$X$ being the antisymmetric $(D-1) \times (D-1)$ matrix with ones on its
upper triangular part.
Then $R_1$ is a rotation of $\alpha$ degrees about the axis $e_1$ and 
$R(\alpha)=W_e^t R_1 W_e$ has $e$ as its axis of rotation. 
\section{Simulations and discussion}\label{s.simulations}
For the simulations the channel 
was modeled for $\nt=2$ transmit antennas and  $\L=2$ taps and for
simplicity at the receiver maximum likelihood decoding has been performed. 
We constructed space time codes with block length $\T=4$, corresponding to
lattice dimension $D=12$, compare \eqref{e.D}. In order to  
obtain full (frequency) diversity for a scenario with $\L=2$ channel taps
these codes are designed for systems with $\K=\L\T=8$ subcarriers.
As a reference, an Alamouti based space time/frequency code of comparable
rate has been simulated. Since the block length of the (space time)
Alamouti scheme is $2$ this code family is well adapted to MIMO-OFDM
scenarios with $\K=2\L=4$ subcarriers. Other choices 
of $\K$ (smaller) would destroy diversity or (larger) add redundancy, which
lowers the rate and would not give fair comparisons. 
Note that the restriction on $\K$ for the Alamouti scheme underlines the
need for new coding schemes, which are 
capable of meeting more general system requirements $(\K,\L,\nt)$. The
spherical space frequency codes proposed in this work achieve this by
appropriately choosing the spherical dimension $D$.

Application of Algorithm \ref{alg.endecoding} produced (by the prescription
of the spherical minimal distance) two sets of spherical space frequency
codes. The first  
set consists of very high rate codes based on the lattices $K_{12}$,
$D_{12}$, $K_{12}$ again, and $A_{12}$ with (space frequency) rates 
$R=1.95, 1.71, 1.65, 1.64$. They will be compared to the 8-PSK modulated
Alamouti scheme of (space frequency) rate $R=1.5$. 
The second set consists of medium rate codes based on the lattices
$K_{12}$, $Z_{12}$, and $A_{12}$ with rates
$R=1.32, 1.15, 1.04$ which are compared to the QPSK Alamouti scheme of
(space frequency) rate $R=1$. 
All these lattice have been rotated as described in the previous section by
an angle $\alpha=\nicefrac{\pi}{2}$ for a better diversity gain. 
Figure \ref{fig.sfcs} shows the simulation results.
%
\begin{figure}[htb]
\begin{center}
   \epsfig{file=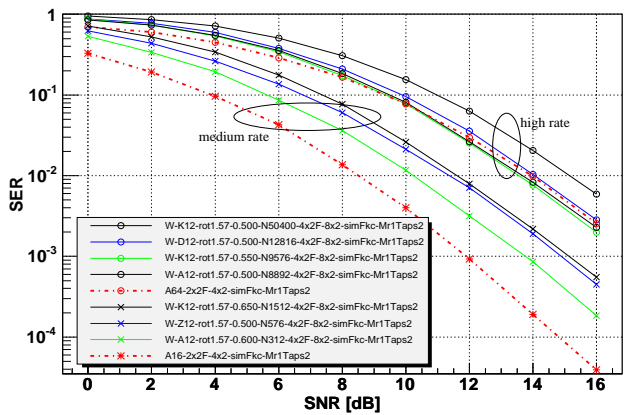,width=1.05\linewidth,height=6cm}
\end{center}
   \caption{Symbol error rate for spherical space frequency codes compared to
     corresponding Alamouti (8-PSK,QPSK) schemes\label{fig.sfcs}}
\end{figure}

The simulations demonstrate that in the high rate regime the spherical space
frequency codes outperform the corresponding Alamouti scheme. Moreover
the Alamouti code has the lowest (space frequency) rate in this comparison. 

For medium rates the picture changes. Here the Alamouti scheme (although
the code with the lowest rate in the simulation) still outperforms the new
spherical space frequency codes. Nevertheless, due to the lattice rotation
all codes seem to have full diversity, even the qubic $Z_{12}$ lattice
performs not too bad. 
\section{Conclusions}
In this work a class of variable rate, fully diverse spherical space
frequency codes based on the scheme \cite{boe.bor.pau}
has been proposed, which applies to arbitrary channel parameters,
i.e. no. of subcarriers $\K$, no. of channel taps $\L$, no. of transmit
antennas $\nt$. The possible rates are determined by the choice of the 
underlying lattice and its designed minimal distance (obtained by an
appropriate scaling of the lattice). This makes the
constructions rather flexible to meet a wide range of
applications. In particular very high rate codes (up to $R=1.95$ for an
$8\times2$ space frequency code in the examples) can be constructed in this
way. The diversity order can be increased by some lattice rotation
and in the examples full diversity has been achieved. The multipath
diversity controlled by the matrix $A$ \eqref{e.standardA} has been chosen
with respect to an advantageous effect on the power distribution among the
subcarriers. 
The simulations demonstrate satisfactory performance results (at least in
the high rate regime), and the lattice based code design offers the
possibility of moderate decoding complexity. 

However improvements 
of the code design and its complexity seem to be possible by careful
analysis of the encoding 
algorithm. The connection of space time/frequency codes with spherical
codes/WBE sequences in CDMA systems comes as a surprise, made possible by a
differential geometric analysis for a (first order) sphere packing approach
to space time/frequency coding. This offers a new direction in
future research efforts.
%
%
%
%
%
%
%
%
%
%
%
%
%
\bibliography{../../../../../refhhi}
\end{document}
%
%
%
%
%
